# Did circoviruses intermediate the recombination between bat and pangolin coronaviruses, yielding SARS-CoV-2?

**Nabil Abid[1,2]\*, Giovanni Chillemi[3,4], Ahmed Rebai[5].**

[1]Laboratory of Transmissible Diseases and Biological Active Substances LR99ES27, Faculty of Pharmacy, University of Monastir, Rue Ibn Sina, 5000, Monastir, Tunisia.

[2]High Institute of Biotechnology of Sidi Thabet, Department of Biotechnology, University of Manouba, BP-66, 2020, Ariana-Tunis, Tunisia. nabil.abid@isbst.uma.tn

[3]Department for Innovation in Biological, Agro-food and Forest systems, DIBAF, University of Tuscia, via S. Camillo de Lellis s.n.c., 01100 Viterbo, Italy. gchillemi@unitus.it

[4]Institute of Biomembranes, Bioenergetics and Molecular Biotechnologies, IBIOM, CNR, Via Giovanni Amendola, 122/O, 70126, Bari, Italy.

[5]Laboratory of Molecular and Cellular Screening Processes, Centre of Biotechnology of Sfax, University of Sfax, PoBox 1177, 3018, Sfax, Tunisia. ahmed.rebai@cbs.rnrt.tn

\**Correspondence to*: Dr Nabil Abid. High Institute of Biotechnology of Sidi Thabet, Department of Biotechnology, University of Manouba, BP-66, 2020, Ariana-Tunis, Tunisia. nabil.abid@isbst.uma.tn

**Abstract:** Since the first reports of a coronavirus (CoV) disease 2019 (COVID-19) caused by severe acute respiratory syndrome virus (SARS-CoV-2) in Wuhan, Hubei province, China, scientists are working around the clock to find sound answers to the issue of its origin. While the number of scientific articles on SARS-CoV-2 is increasing, there are still many gaps as to its origin. All studies failed to find a coronavirus in other animals that is more similar to human SARS-COV-2 than the bat virus, considered to be the primary reservoir. In this paper we address a new hypothesis, based on a possible recombination between a DNA and SARS-CoV viruses, to explain the rise of SARS-CoV-2. By comparing SARS-CoV-2 and related CoVs with circoviruses (CVs), we found strong sequence similarity of the genomic region at the 3'-end of Bat-CoV ORF1a and the origin of replication (Ori) of porcine CV type 2 (PCV2), as well as similar RNA secondary structures of the region encompassing the cleavage site of CoV S gene with the PCV2 Ori. This constitutes a primary evidence that supports a possible recombination, which occurrence might explain the origin of SARS-CoV-2.

**1. Introduction**

Many theories about the origin of the SARS-CoV-2 have been proposed; the most debated ones have been: its natural emergence after passing from bats to an intermediate animal, which served as a springboard for SARS-CoV-2 to jump from bats to humans *(1)*, and the virus being deliberately engineered and accidentally released by humans *(2)*. The hypothesis of natural emergence is far more supported by recent data and two scenarios have been suggested (3); the first is natural selection in an animal host before zoonotic transfer which necessitates an animal intermediate that have a high population density (to allow natural selection) and an ACE2-encoding gene similar to the human ortholog, since the SARS-CoV-2 virus have acquired both the polybasic cleavage site and mutations in the Spike protein suitable for human ACE2 binding. The second scenario is natural selection in humans following zoonotic transfer, where the progenitor of SARS-CoV-2 jumped into humans and acquired new genomic features through adaptation during undetected human to human transmission.

It is now recognized that the newly emerged SARS-CoV-2 Spike (S) glycoprotein gained specific features that facilitated its spread as compared to its closely related Bat-CoV RaTG13 strain, as supported by both phylogenetic analysis *(4)* and a single nsp8 gene, typical of Bat-CoVs *(5,6)*. The most remarkable event in SARS-CoV-2 S glycoprotein is the insertion of a cleavage site (residues PRRA) at the boundary between the $S_1$ and $S_2$ subunits, compared to Bat-CoV RaTG13 strain (Gisaid ID: EPI_ISL_402131). The severity of infection has not been yet fully linked to the newly acquired furin-like cleavage site, however has likely facilitated the host species jump and a dramatic increase in cell-cell fusion capacities. Such events were previously reported for Sendai virus following the insertion of a second furin cleavage site in the F protein *(7)*.

Additionally, the insertion of a furin-like cleavage site in SARS-CoV-2 S glycoprotein is reminiscent of a low-pathogenic avian influenza virus (AIV) *(8,9)*, which, when introduced into a poultry farm, acquired a polybasic cleavage motif that caused a deadly outbreak of highly pathogenic virus. However, several questions are still unanswered concerning the reason why the acquisition of the cleavage site is restricted to a particular genome region, as reported previously for AIV *(10)* and whether SARS-CoV-2 is a "Chimera" of two viral strains or if it resulted from the accumulation of point mutations, as hypothesized in most published studies (1). CoVs are characterized by the reduced error rate of their viral polymerases, much lower than that of other RNA viruses *(11,12)*, unless they recombine their genomes with a related CoV. This feature, that has been observed for SARS-CoV-2 infection (as estimated from public genome sequences released since December 2019), might ultimately limit the mutagenic variability of the virus over a short period of time. Therefore, recombination might prove to be a prominent driving factor in the evolution and the expansion of species range and cellular tropism.

Recent studies reported the detection of a Bat-CoV RmYN02 strain (Gisaid ID: EPI_ISL_412977) containing insertions in the $S_1/S_2$ cleavage site *(13)* and the detection of two human SARS-CoV-2 variants (SARS-CoV-2 Variants 1 and 2) *(14)* showing deletion mutations in the furin-like cleavage site and its flanking sites. More recent study reported three human variants (referred here as SARS-CoV-2_mut1, SARS-CoV-2_mut2, and SARS-CoV-2_mut3) showing deletions at the $S_1/S_2$ junction *(15)*. The detection of the human variants, *in vivo* and/or *in vitro*, suggests that this region of the S gene is under strong selective pressure, given that replication in permissive cells leads to the loss of this adaptive function.

Given the importance of this region, we decided to further analyze a possible origin of the furin-like cleavage site in the SARS-CoV-2 S gene.

The most important question is whether these insertions/deletions are due to homologous recombination or might have resulted from a recombination between two different viruses. Here, we consider the hypothesis of a possible recombination between RNA and DNA viruses to explain the emergence of SARS-CoV-2. Although the recombination between RNA and DNA viruses was considered, for a long time, as an unusual and extremely rare event, it was previously described by the isolation of a circovirus-like genome, called Boiling Springs Lake RNA-DNA hybrid virus (BSL-RDHV) from an acidic hot lake *(16,17)*. It was reported that the gene for the rolling-circle replication initiation protein (RC-Rep) of BSL-RDHV is inherited from a circovirus-like ancestor whereas the capsid protein (Cap) gene is most closely related to that of ssRNA viruses *(16)*. Additionally, a metagenomic study tracing the prevalence of circular DNA viruses in tissue specimens and environment samples, followed by sequence annotation using bioinformatics tools, reported recombination events, never described before, between ssDNA viruses and ssRNA viruses (18).

## 2. Material and Methods

### 2.1. Genetic data and sequence alignment

Publicly available genomic CoVs sequences were obtained from GenBank (https://www.ncbi.nlm.nih.gov/) and GISAID (https://www.gisaid.org/) (as by 08/09/2020). We used basic local alignment search tool BLAST (National Center for Biotechnology Information [NCBI], Bethesda, United States) *(19)*. The use of local alignment is justified by the fact that the CVs and CoVs are too divergent to retain discernable sequence similarity using global alignment algorithms. Nucleotide BLAST was used with appropriate parameters (Word size: 7; Gap Existance:4, Gap Extension: 2). Data mining was carried out using local database comprising a metagenomic data of DNA viruses in animal specimens and environmental samples, reported previously *(18)*. To evaluate the sequence variability, genome sequences were aligned using ClustalW *(20)* and MAFFT *(21)* and refined manually. Alignments for region flanking the cleavage site of the S glycoprotein and the PCV2 Ori were extracted from the genome alignment.

### 2.2. Secondary structure prediction

The centroid secondary structures (SS) were first generated by RNAfold (http://rna.tbi.univie.ac.at/) using default parameters *(22)*, then visualized and edited using VARNA v.3-93 *(23)*. In order to show SS variability, the missing sequence motifs in viral strains (RaTG13 and Pangolin CoVs) were completed by their corresponding sequences in SARS-CoV-2.

## 3. Results

### 3.1. Sequence homology and secondary structure analyses

We selected, as a query for sequence homology search (using nucleotide BLAST), a mosaic region comprising the insertion region of the recently detected Bat-CoV RmYN02 strain (Gisaid ID: EPI_ISL_412977) and its missing upstream and downstream nucleotide sequences from SARS-CoV-2 Wuhan strain (Genbank accession nº NC_045512) (Fig. 1). The similarity search was carried out against circoviruses (taxid:39725) with appropriate alignment parameters (see material and methods). Three similar 11-nt sequences were shown in the genomes of PCV2 (Genbank accession nº MF326358 and MK377563) and Beak and feather disease virus (Genbank accession nº MN175611) (Fig. 1). The first two similar regions of PCV2 correspond to the junction region between Cap and replication initiator protein (Rep) genes, known to be the DNA origin of replication (Ori) of PCV2 *(24)*, while the third one corresponds to a reverse complement region in the Cap gene of Beak and feather disease virus. The main structural feature of the Ori is its stem-loop secondary structure (SS). Based on this result, we carried out the alignment of the region flanking the furin-like cleavage site of SARS-CoV-2, SARS-CoV-1, and their related strains.

Variants 1 and 2 of SARS-CoV-2 (from the original article) *(14)*, human SARS-CoV-2 Wuhan strain (Genbank accession nº NC_045512; position from 23554 to 23639), the three SARS-CoV-2 mutants described above (SARS-CoV-2_mut1, SARS-CoV-2_mut2, and SARS-CoV-2_mut3; from the original article), Bat-CoV RaTG13 strain (Genbank accession nº MN996532; position from 23536 to 23609), two Human SARS-CoV-1 (Genbank accession nº DQ182595; position from 23425 to 23494; Genbank accession nº DQ640652; position from 23447 to 23516 ), Bat-CoV RmYN02 strain (Gisaid ID: EPI_ISL_412977; position from 23415 to 23482), Pangolin-CoV isolate MP789 (Genbank accession nº MT121216; position from 23400 to 23473), Bat_CoV isolate Rs4874 (Genbank accession nº KY417150; position from 23442 to 23511), and Civet related CoV-1 (Genbank accession nº AY572034; position from 23415 to 23484 ) were used in the present study to generate a multiple sequence alignment (Fig. 2A). Interestingly, data mining of unclassified CVs isolates showed similarity of a unexpected Rep 3'-5' sequence region of a Raccoon CV (isolate ctcc28) with both SARS-CoV-1 and SARS-CoV-2 S genes; it was then added to the sequence alignment (Genbank accession nº MK012475; position from 1690 to 1763).

The results showed a high variability, mainly in the downstream region of the alignment (Fig. 2A). The used Raccoon CV showed sequence similarity with SARS-CoV-1 and its related strains (yellow boxes), mapping mostly to the beginning of the alignment and to a region specific of SARS-CoV-1 strains. However, the highest sequence similarity of Raccoon CV was shown with SARS-CoV-2 and its related strains, mainly in the high variable region (blue boxes).

To assess whether S gene of CoVs shows similar SS structure as that of CV, we carried out the SS analysis of the RNA region flanking the cleavage site of different CoVs (Human SARS-CoV-2 Wuhan strain, Bat-CoV RaTG13 strain, and Pangolin-CoV isolate MP789) and compared them with the well-known SS of the PCVs Ori *(24,25)* (Fig. 3).

The results showed five similar nucleotides at the loop structure (5'-gUAUA-3') and four similar nucleotides (5'-gCgC-3') in the stem of PCV2 and SARS-CoV-2 SS palindrome (Fig. 3). While these nucleotides are totally or partially shared by Bat-CoV RaTG13 strain and Pangolin-CoV isolate MP789, they were different in Bat-CoV RmYN02 strain and SARS-CoV-1 (data not shown). However, nucleotide substitutions at specific positions downstream the the cleavage site of Bat-CoV RmYN02 allow the generation of similar SS with SARS-CoV-2 (data not shown), suggesting that nucleotide mutations at the stem loop structure of CoVs might hamper the formation of similar SS as PCV2.

As reported previously, the SS of PCV2 Ori showed four hexa-nucleotide sequences (H1-4) and two penta-nucleotide sequences (P1 and P2) *(25)*, whereas we report here that SARS-CoV-2 showed three H (H1-3) and one P; both viruses shared the H3 and partially P1, generating the furin-like cleavage site. The Bat-CoV RatG13 strain and Pangolin-CoV isolate MP789 shared two H (H1 and H2) with SARS-CoV-2 (Fig. 3).

According to these SS structures, the Bat-CoV RaTG13 strain is more closely related to SARS-CoV-2. The difference was an insertion of P and H3 sequences, identical to P1 and H3 of PCV2, generating a cleavage site (Fig. 3).

*3.2. Sequence homology analysis of bat-CoV and PCV2 Ori*

Besides the SS homology between CoVs and PCV2, these findings suggest that Bat-CoV/PCV2 recombination occurs only when both viruses have identical nucleotide sequences at the region flanking the cleavage site of CoVs and the PCV2 Ori. Thus, the region encompassing H1, H2, P1, and H3 (5'-CggCAGCggCAgCACCTCggCgg-3') of the PCV2 Ori was used as query for similarity search among CoV strains (Fig. 2B). The results showed sequence similarity with the 3'-end of ORF1a of three recently published Bat-CoV strains: BtRt-BetaCoV_1 (Genbank accession nº MT337387; position from 12052 to 12127), BtRt-BetaCoV_2 (Genbank accession nº MT337386; position from 12046 to 12121), and BtRt-BetaCoV_3 (Genbank accession nº MT337385; position from 11963 to 12038), compared to other SARS-CoVs (Fig. 2B). Surprisingly, these sequences are mapped to the 3'-end of ORF1a of these Bat-CoVs, while we expected them in the S gene, as a motif in the query sequence unique to the cleavage site of SARS-CoV-2 S gene (5'-CggCgg-3') is shared by PCV2 Ori.

During DNA replication of PCV2, the viral Rep protein of PCV2 nicks an octa-nucleotide sequence (5'-AgTATT$^\dagger$AC-3') of the Ori (loop structure) between $T_6$ and $A_7$ to generate a free 3'-OH end for initiation of plus-strand DNA replication *(24)*. Interestingly, these Bat-CV strains have a similar nucleotide sequence, mainly at the nick site (5'-AgTAgTT$^\dagger$AC-3') (Fig. 2B). The SS prediction of the region of CoVs ORF1a did not show similar structures with PCV2 Ori (data not shown), otherwise we could postulate a recombination event between the two viruses driven by template switching of the polymerase during the virus replication, given the high variability among CoV strains at the downstream of stem-loop structure. More sampling within bat CoV population may help to better characterize the sequence similarity shown in the present study. These results, taken altogether, suggest that two separate events might have occurred. While the first analysis showed sequence similarity between the 3'-end of ORF1a of several Bat-CoVs and PCV2 Ori, the second showed sequence similarity between a unclassified Rep CV sequence from the skin of Raccoon and SARS-CoVs as well as SS match of the region encompassing the cleavage site of a Bat-CoV RaTG13, Pangolin CoV, and Human SARS-CoV-2 strains with PCV2 Ori. Interestingly, the furin-like cleavage site (5'-CCTCggCgg-3') was shared between PCV2 Ori and Raccoon CV Rep gene. Although the transmission of PCVs to Raccoon population is recently reported (*26*), the exact mechanism(s) underlying the genetic rearrangement within CV genomes is still unknown.

These *in silico* findings are far to be a definitive answer on the SARS-CoV-2 origin but we think they can be a key piece in the Wuhan puzzle. In fact, our assumption postulates that CVs (Porcine and/or Raccoon) likely played, in the course of evolution, a role as an intermediate between Bat/Pangolin CoVs due to sequence similarity and SS match with the region of the 3'-end of ORF1a and the region flanking the cleavage site of the S gene, respectively. The genetic rearrangement and recombination may enhance the acquisition of the cleavage site in the S gene, resulting in a rapid evolution of the virus in Bats and its jump into humans (Fig. 4).

## 4. Discussion

Some events in the past few years about viral infections among animals in China need to be highlighted, supporting a bat-swine transmission of CoV infection in China.

In 2017, it was reported that SARS-Like-CoVs in bats were discovered in a cave in Yunnan province, China, while tracing the SARS-CoV pandemic in 2003 (SARS-CoV-2003). The researchers declare that these viruses used the ACE2 receptor to infect cells and could replicate efficiently in primary human airway cells *(27)*. This study revealed that some of these SARS-Like-CoVs circulating in this cave are highly diverse in the S gene whereas other strains with high genetic similarity to SARS-CoV-2003 in the hypervariable N-terminal domain (NTD) and receptor-binding domain (RBD) of the $S_1$ subunit. However, this study did not report the insertion of a furin cleavage site within the sequenced genomes of the new detected SARS-Like-CoVs, suggesting that SARS-CoV-2 strain did not originate directly from those isolates.

In the same year (2017), another new bat-HKU2-like porcine CoVs, named porcine enteric alphacoronavirus (PEAV), was isolated from suckling piglets in Guangdong, China *(28-30)*. The full genome sequence of PEAV GDS04 strain shared high nucleotide identities (≈95%) with the reported bat-HKU2 strains (EF203064). One year later, another novel swine acute diarrhea syndrome coronavirus (SADS-CoV-2018) was reported in Fujian, China *(31)*. This virus was shown to be genetically related to previously detected PEAV GDS04 and bat-CoV-HKU2 strains with 99.6% nt and 98.9% nt identity, respectively. Furthermore, the evolutionary analysis showed that SADS-CoV-2018 was more similar to bat-origin SADS-related CoVs (SADSr-CoVs)*(32)* than to those of swine-origin SADS-CoVs, suggesting that SADS-CoV-2018 might have originated from bats.

The diversity and prevalence of environmental RNA and DNA viruses as epicenters of uncommon recombination events are still very poorly studied *(33)*. Although the DNA/RNA recombination was not shown previously between PCV2 and Bat-CoV, heterologous inter-family recombination event was reported between Bat-CoV and bat orthoreovirus *(34)*.

The sequence similarity and SS match could constitute another mechanism by which the recombination events occur between these two unrelated viruses. While the SS of CVs Ori was discussed in the present study as the main structure for the initiation of virus replication, these structures regulate many stages of the viral replication cycle of RNA viruses, including genome replication and packaging, intracellular trafficking *(35)*, and play a potential role in genetic recombination *(36-39)*.

The high variability and rearrangement of the region downstream the stem-loop SS were reported previously for PCV1 *(40)*, closely related to PCV2, to yield deletion and/or extensive nucleotide reorganization of the hexa-nucleotide sequences. In particular, mutations engineered into H1/H2 of nonpathogenic PCV1 were invariably deleted so that the downstream H3/H4 was placed next to the palindrome *(40)*. According to this same study, viral genomes with mutations engineered into both H1/H2 and H3/H4 underwent extensive nucleotide reorganization to yield progeny viruses containing either H3/H4, h-like/H4, or h-like/H3/H4 sequences, generating sequences in a similar way as the downstream region of SS of Bat-CoV RmYN02 and SARS-CoV-1 strains (data not shown).

DNA replication of CVs is initiated by cleaving the loop structure at a specific nick site using the synthesized viral Rep protein *(41)*. Therefore, we suggest that the deletions in the cleavage site in recently detected SARS-CoV-2 variants, even though they showed similar SS, are a result of a selective pressure on the adaptive sequence rather than specific point mutations or homologous recombination, in a similar way as for CVs.

Finally, it is worth noting that the sequence mapped to the PCV2 Ori was reported to have a CpG oligodeoxynucleotide (ODN) *(42)* and therefore similar CoV sequence mapping to that position could modulate the immune response. This needs further experimental investigation.

## 4. Conclusion

The RNA viruses as new potential candidate agents for the global pandemics were extensively discussed *(43)*. However, due to their high genetic variability, inter- and intra-species recombination strategies increasing the already huge diversity, it is impossible to anticipate the emergence of a new viral strain. The emergence of SARS-CoV-2 could be seen as an opportunity to elucidate some uncommon RNA/DNA recombination events. We report in the present study a new perspective concerning the emergence of SARS-CoV-2 by focusing on the cleavage site of the S gene and similar events that might have occurred in the genome of SARS-CoV-2. This suggests the importance of monitoring SARS-like CoVs in Bats and CVs in animal herds as well as the establishment of effective strategies to hamper interspecies spread of these two viruses.

**Acknowledgments:** This work was supported by the Tunisian Ministry of Higher Education and Scientific Research, the 'Departments of Excellence-2018' Program (Dipartimenti di Eccellenza) of the Italian Ministry of Education, University and Research, DIBAF-Department of University of Tuscia, and the Project 'Landscape 4.0–food, wellbeing and environment'.

**Author contributions:** Conceptualization, N.A.; methodology, N.A. and G.C.; software, N.A.; writing—original draft preparation, N.A., G.C., and A.R.; writing—review and editing, N.A., G.C and A.R.

**Competing interests:** Authors declare no competing interests.

**Data and materials availability:** All data is available in the main text.

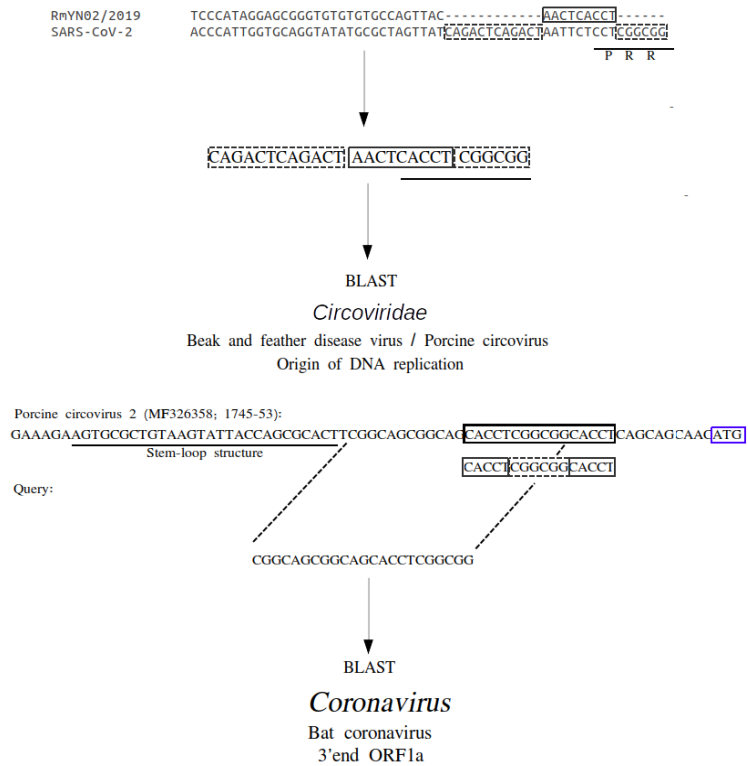

**Figure. 1**. Schematic diagram of the two steps used to compare the query sequence. The first step was carried out using mosaic region as query for SARS-CoV-2 and Bat-CoV RmYN02 strain against the available circovirus strains. The used sequences were shown by dashed and line rectangles and the stem-loop sequence was underlined. The start codon for PCV2 Rep gene was shown by blue box. The second step was carried out using the PCV2 Ori region against the available coronavirus strains.

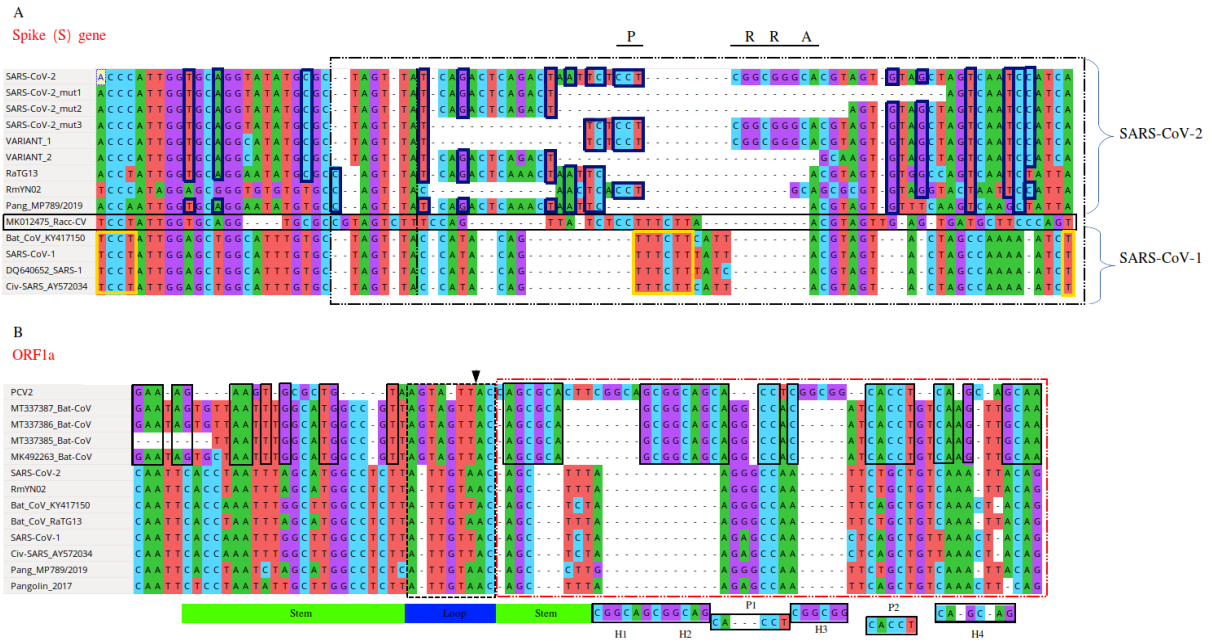

**Figure. 2**. Schematic representation of the sequence alignments. (A) The alignment of the region of the S gene cleavage site and its flanking region of SARS-CoV-1, SARS-CoV-2, and their related strains. The Raccoon CV Rep sequence is highlighted by a black box. The identical nucleotide positions of Raccoon CV with SARS-CoV-1 and its related strains are shown by yellow boxes whereas blue boxes highlighted the identical nucleotide positions with SARS-CoV-2 and its related strains. The 2 dots 3 dashes black box is used to highlight the variable region of the sequence alignment. The one-letter amino acids of the acquired cleavage site are shown by horizontal lines. (B) The alignment of the PCV2 Ori with 3'-end ORF1a of CoVs. Identical nucleotide positions between PCV2 Ori and Bat-CoVs are shown by black boxes. The 2 dots 3 dashes red box is used to highlight the variable region of the sequence alignment. The stem-loop, hexanucleotide (H), and pentanucleotide (P) sequences are shown by panels underneath the figure. The nick site is indicated by an arrow.

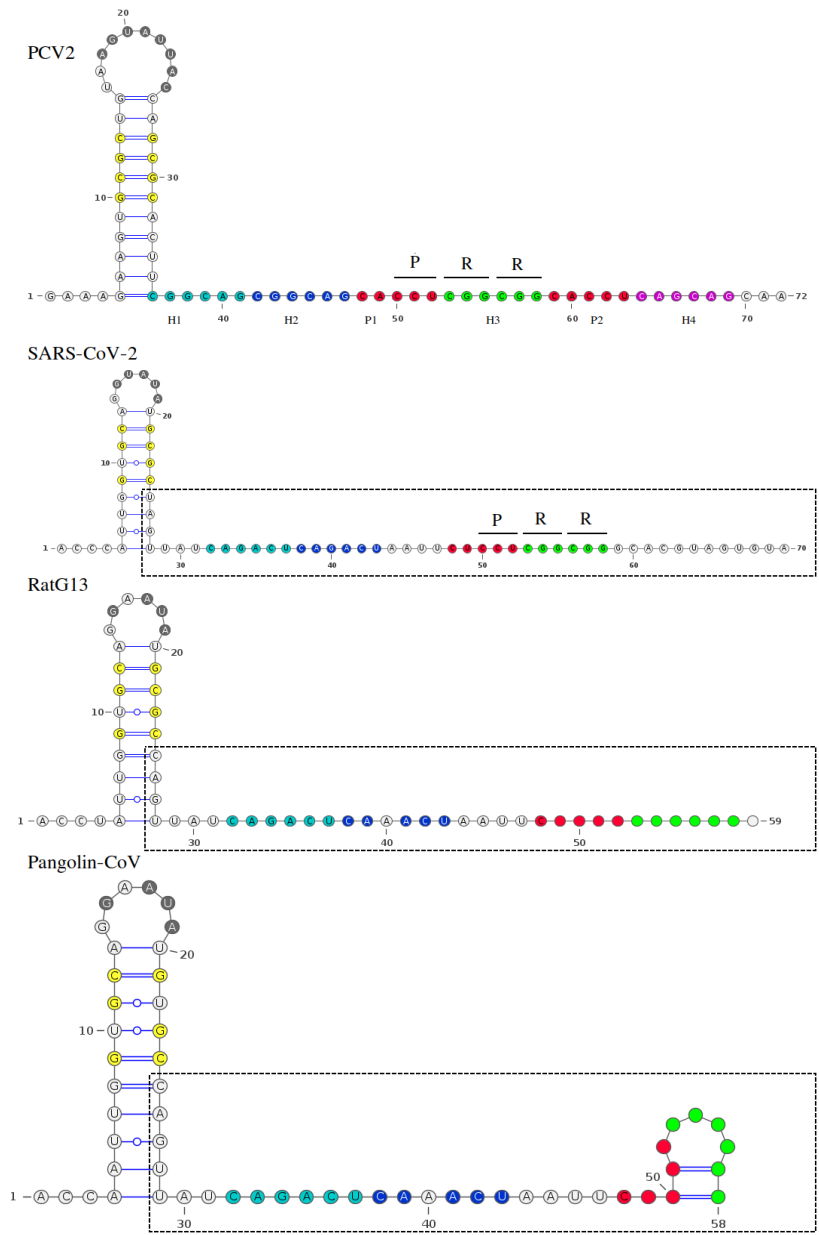

**Figure 3**. Secondary structure comparison of different virus strains generated by RNAfold and edited by VARNA v.3-93. Similar nucleotides in the stem-loop between PCV2 and CoVs (human, bat, and pangolin) are shown by yellow and grey, respectively. The hexanucleotides (H) and pentanucleotides (P) are shown by different colors. The variable region in the alignments (Fig.2) is highlighted by ultrafine dashed boxes. The one-letter amino acids of the cleavage site are shown for SARS-CoV-2.

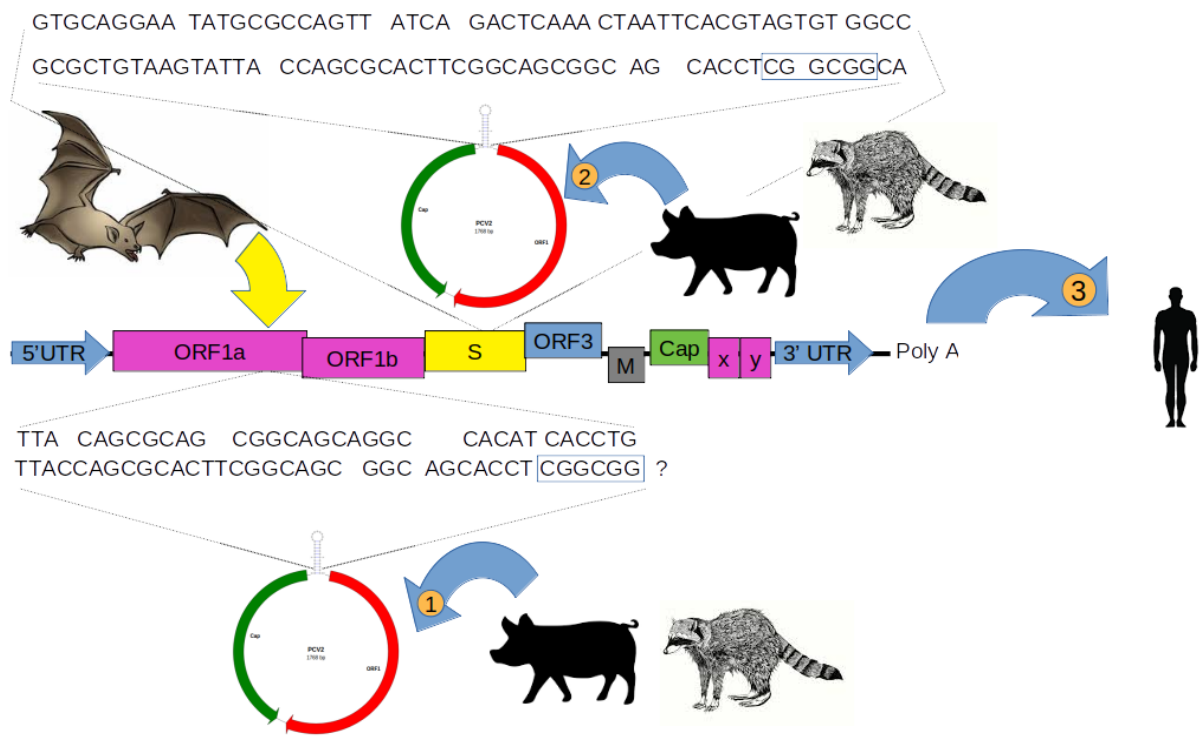

**Figure 4.** Genome organization of PCV2 and Bat-CoV (Genbank accession nº NC_048212). The sequences of the PCV2 Ori and regions of the S glycoprotein encompassing the cleavage site as well as region showing sequence similarity with PCV2 Ori are illustrated with dashed lines. Numbers in arrows indicate the suggested steps for genetic recombination between Bat-CoV and animal CVs. The remarkable insert in the SARS-CoV-2 S glycoprotein is indicated by a blue box and a question mark. The bat, swine, raccoon, and human are used to show the host species involved in the present study.